# Dynamic Behavior and Microstructural Properties of Cancellous Bone.

S. Laporte[1], F. David[1], V. Bousson[2] and S. Pattofatto[3]

[1] Arts et Métiers ParisTech, CNRS, LBM, 151 Boulevard de l'Hôpital 75013 Paris, France
[2] Service de Radiologie Ostéo-Articulaire, Hôpital Lariboisière, Paris, France
[3] LMT-Cachan (ENS Cachan/CNRS/Université Paris 6/ PRES UniverSud Paris) 61 av. du Président Wilson, F-94230 Cachan, France

**Abstract.** The aim of the presented study is to identify some properties of the dynamic behavior of the cancellous bone and to identify the link between this mechanical behavior and the microstructural properties. 7 cylinders of bovine cancellous bone (diameter 41 mm, thickness 14 mm) were tested in quasi static loading ($0.001$ $s^{-1}$), 8 in dynamic loading ($1000$ $s^{-1}$) and 10 in dynamic loading ($1500$ $s^{-1}$) with a confinement system. All the specimens were submitted to imaging before the tests (pQCT) in order to indentify two microstructural properties: Bone Volume / Total Volume – BV/TV – and Trabeculae Thickness – Tb.Th. The behavior of bovine cancellous bone under compression exhibits a foam-type behavior over the whole range of strain rates explored in this study. The results show that for the quasi-static tests only the stresses are correlated with BV/TV. For the unconfined dynamic tests, the yield stress is correlated to BV/TV and the plateau stress to BV/TV and Tb.Th. For the confined tests, only the plateau stress is correlated to BV/TV and Tb.Th. The effect of strain rate is an increase of the yield stress and the plateau stress. The confinement has an effect on the measured values of compression stresses that confirms the importance of marrow flow in the overall behavior.

## 1. Introduction

Nowadays, the numerical modeling of the human body is one of the prior interests in the car industry in order to evaluate the injury risk in case of an accident [1,2]. To construct these models, it is necessary to implement relevant behavior laws in finite elements code and to identify the specific parameters for each human tissue in a difficult context: properties are widely scattered due to inter individual differences.

There are two different types of bone in the human skeleton: the compact bone and the cancellous bone. The cancellous bone is a complex network of plates and beams (trabecular bone) filled with bone marrow. Mechanical properties of bones and their links with medical imaging are widely studied for quasi static loading in the field of osteoporosis [3]. Precedent studies have shown that the cancellous bone has a mechanical behavior close to cellular materials one [4,5,6,7]. Furthermore, the mechanical properties are dependent of the loading: traction and compression behavior are different [8], and the flow of the bone marrow has an influence on the apparent mechanical behavior of the cancellous bone [9]. However there is a lack of knowledge for the high dynamic properties of this type of bone.

The aim of the presented study is then to characterize the mechanical properties of the cancellous bone for compression loading at low and high strain rates and to identify the links between these mechanical properties and the microstructural description of the material. The proposed experiments are conducted on bovine cancellous bone.

## 2. Material and Methods

### 2.1 Samples

A total of 15 distal parts of bovine femoral bones were used for this study (72 hours *post mortem*) and frozen at -20°C. Two cylinders of cancellous bone were then harvested from each frozen bone in the sagittal plane (diameter 41 mm and thickness 14 mm). Specimens were slowly thawed 12 hours at +5 °C before the tests.

### 2.2 Microstructure Properties Identification

Each frozen cylinder was scanned using the peripherical quantitative tomodensitometry technique (pQCT). A specific setting was developed to adapt this *in vitro* measurement to the XtremeCT (Scano Medical, spatial resolution: 4.2 pl.mm$^{-1}$, voxel size: 42 μm). Two descriptors of the microstructure of the cancellous bone were used in this study: the ratio of bone volume relatively to the total volume of the specimen BV/TV and the mean thickness of the trabeculae Tb.Th.

### 2.3 Mechanical Tests – Mechanical Properties Identifications

Three different compression tests were performed in the presented study: 7 tests in quasi static loading (0.001 s$^{-1}$), 8 in dynamic loading (ca. 1000 s$^{-1}$) and 10 in dynamic loading (ca. 1500 s$^{-1}$) with a confinement system.

Quasi Static tests were performed with an Instron Static 5500R 1185 device. A prescribed compressive displacement was applied to the bone specimen giving a constant nominal strain rate of 0.001 s$^{-1}$. Dynamic and confined dynamic tests were performed by dynamic compression tests on Split Hopkinson Pressure Bars (SHPB). The nominal strain rates were of ca. 1000 s$^{-1}$ for the compression tests and ca. 1500 s$^{-1}$ for the confined ones. The apparatus is made of 40 mm diameter nylon bars. Input and output bars are 3 m long and the striker is 1 m long. The striker is launched with a gas gun at an impact velocity about 15 m/s. A fast camera is also used to follow qualitatively the deformation of the specimen during compression [10]. The pictures are 128 x 272 px and the frame rate is 37'500 fps. Confined dynamic tests were done by placing the specimen into a confining cell to limit the expansion of the trabecuale structure and the flow of the bone marrow.

For all tests, the results are given as nominal stress versus nominal strain for different strain rates. The subsequent identification is focused on three mechanical parameters classical for a cellular material submitted to a compression loading: the apparent Young's modulus E, the yield stress $\sigma_y$ and the plateau stress $\sigma_p$. For dynamic tests performed on SHPB device, the Young's modulus is taken from an elastic simulation procedure usually used to calibrate the time-origin of the waves [11].

### 2.4 Statistical Indicators

The Kruskal-Wallis statistical test was performed in order to verify that there was no difference between the 3 lots of specimens in terms of microstructure. The Kolmogorov-Smirnov statistical test was performed to analyze the significant differences of the mechanical properties for the different boundary condition. Finally, the correlations between mechanical and microstructure properties were analyzed using the Spearman test. The test significance was set for a p-value less than 5 %.

## 3. Results and Discussion

### 3.1 Microstructure Properties

Figure 1 presents the 3D reconstruction of a half specimen used to calculate the microstructure properties. A total of 5 specimens were eliminated because they exhibited a non-homogeneous structure: 3 for the quasi static tests and 2 for the dynamic ones. Even if these specimens were tested, the results were not included.

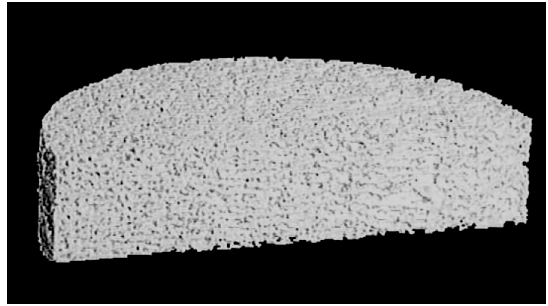

**Figure 1.** pQCT 3D reconstruction of a half specimen, diameter 41 mm, only the trabecular bone is reconstructed.

No significant difference was found between the 3 lots of specimens for BV/TV (p = 0.54) and Tb.Th (p = 0.36). For the tested specimens, the mean value of BV/TV is about 17 % with a standard deviation of 3.9 % and the mean value of Tb.Th is about 106 μm with a standard deviation of 16 μm.

**3.2 Mechanical Properties**

The behavior of bovine cancellous bone under compression is measured as a foam-type behavior over the whole range of strain rates explored in this study (Figure 2a). In this way, the curve exhibits a first linear elastic region, a quasi-constant stress region called the plateau and a densification region, detected only for quasi-static tests in this study due to the small duration of the incident wave during the SHPB experiments. During the unconfined experiments the bone marrow is expulsed as depicted in Figure 2b with images taken during a dynamic test.
As a conclusion, mean values and standard deviations of the different mechanical properties are presented in Table 1. First, we can notice that these values are coherent with the literature [9].

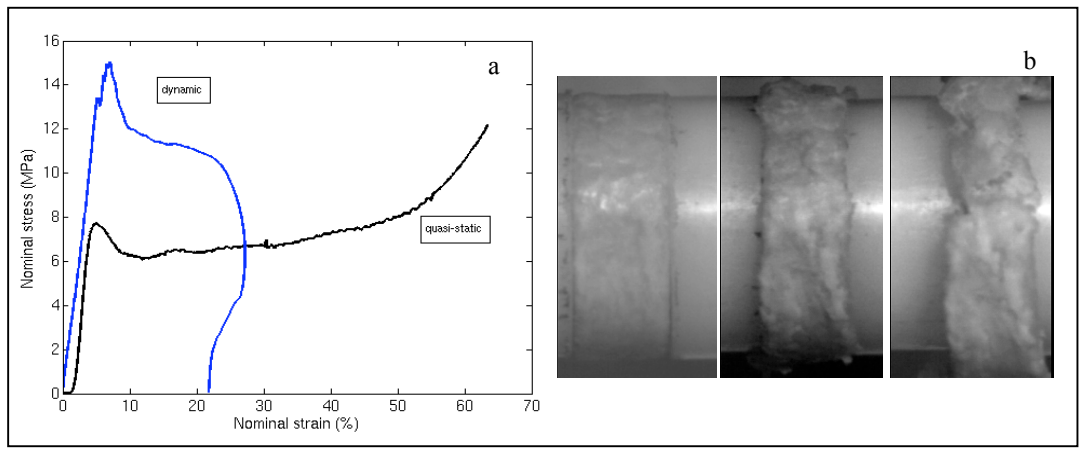

**Figure 2.** a) Quasi static and dynamic compression curves for two specimens that have "close" microstructural parameters. b) Example of dynamic test – expulsion of the bone marrow.

**Table 1.** Mean (bold) and standard deviation (italic) of the mechanical properties for the different tests.

|  | Young's Modulus E *(MPa)* | Yield stress $\sigma_y$ *(MPa)* | Plateau stress $\sigma_p$ *(MPa)* |
|---|---|---|---|
| Quasi-static (0.001 s$^{-1}$) | **269** *(52)* | **5.9** *(1.4)* | **5.5** *(1.1)* |
| Dynamic (1000 s$^{-1}$) | **170** *(77)* | **10.6** *(3.1)* | **8.5** *(3.8)* |
| Confined Dynamic (1500 s$^{-1}$) | **315** *(62)* | **20.4** *(1.4)* | **15.8** *(2.6)* |

The yield stress increases by 80 % and the plateau stress by 55 % with strain rate; the increases due to the confinement are about 250 % for the yield stress and 190 % for the plateau stress. These effects can be explained by the viscoelastic properties of the trabecular bone and bone marrow. The confinement has an effect on the different mechanical properties obtained during the compression tests that confirms obviously the importance of marrow flow in the overall behavior. Thus, for confined experiments, the limitation of bone marrow expansion leads to an increase of the apparent Young's modulus. For the non confined tests, the apparent Young's modulus seems to decrease with strain rate even if this difference is not significant (p = 0.09). This decrease could be explained by local properties of the trabecular bone: variation of mechanical properties at the microstructure level, local buckling of the trabeculae or dynamic effect. Further studies are needed to understand this specific behavior.

**3.3 Influence of the structural properties on the mechanical properties**

The figure 3 presents the variations of the mechanical properties with the structural properties for the 3 different tests.

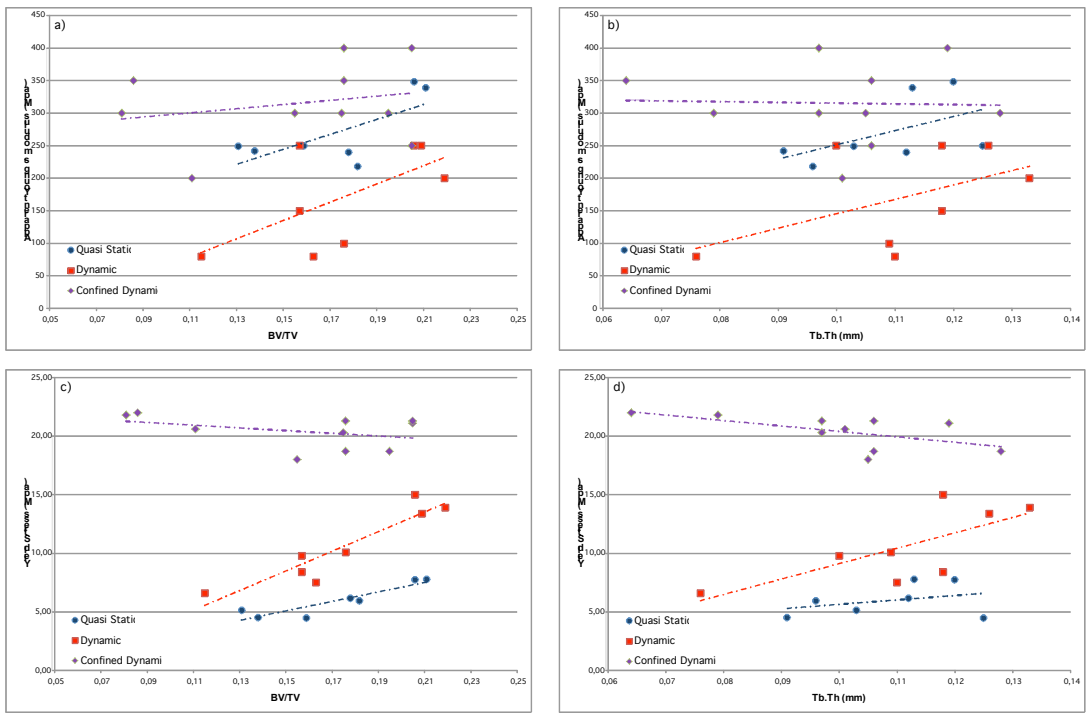

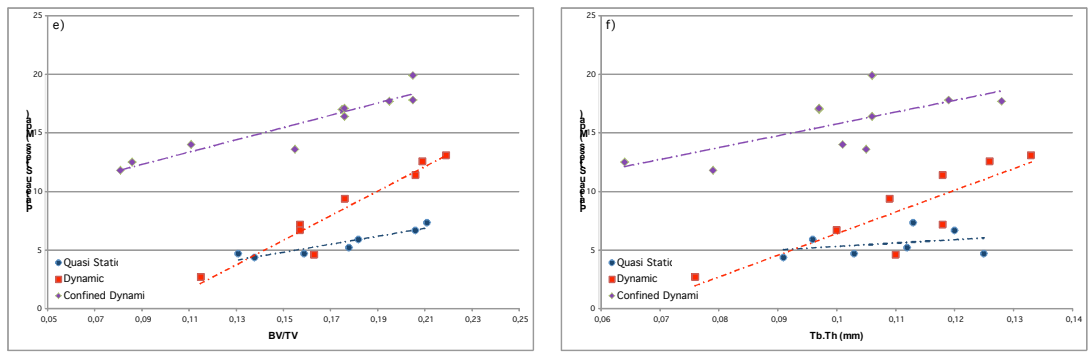

**Figure 3.** Variations of the mechanical properties with the structural properties for the 3 different tests. a) Apparent Young's Modulus E versus ratio bone volume total volume BV/TV, b) E versus mean trabeculae thickness Tb.Th, c) Yield stress $\sigma_y$ versus BV/TV, d) $\sigma$p versus Tb.Th, e) Plateau stress ($\sigma$p) versus BV/TV, f) $\sigma_p$ versus Tb.Th.

For the quasi static and non confined dynamic tests, the apparent Young's modulus E, the yield stress $\sigma_y$ and the plateau stress $\sigma_p$ increase with the ratio bone volume total volume BV/TV and mean thickness of the trabeculae Tb.Th. However, only the stresses have significant correlations with BV/TV: a) $\sigma_y$ versus BV/TV: quasi static p = 0.023, dynamic p = 0.007, b) $\sigma_p$ versus BV/TV: quasi static p = 0.002, dynamic p = 0.001. For dynamic tests, $\sigma_p$ is also correlated with Tb.Th (p = 0.006). These results confirm that the resilience of the cancellous bone increases with the bone volume.
Concerning the confined dynamic tests, E seems to be constant with BV/TV and Tb.Th. This behavior could be explained by the influence of bone marrow, in fact, this test measures the compressive behavior of the bone marrow that represents between 90 % and 75 % of the total volume of the cancellous bone. This property could also explain the slight decrease of $\sigma_y$ with BV/TV and Tb.Th. Finally $\sigma_p$ is correlated with BV/TV (p < 0.0001) and with Tb.Th (p = 0.021). This result underlines the influence of bone during the densification of the specimen.

## 4. Conclusion

In conclusion, the response of the bovine cancellous bone to compression is a foam-type behavior over the whole range of strain rates explored in this study (0.001 s$^{-1}$ and 1000 s$^{-1}$). The effects of strain rate are an increase of the yield and plateau stresses and a slight decrease of the apparent Young's modulus. The bone volume in the cancellous bone has an influence on the yield and plateau stresses for the quasi static and dynamic tests. Only the plateau stress is correlated with the trabeculae thickness for the dynamic and confined dynamic tests. This study underlines the influence of the bone marrow on the global behavior of the cancellous bone. Further studies are needed to explore and to model this specific material.